\documentstyle[12pt,epsf]{article}
\begin{document}

\title{Pionic Degrees of Freedom in Atomic Nuclei and Quasielastic
Knockout of Pions by High-Energy Electrons}

\author{V.G. Neudatchin, N.P. Yudin,
L.L. Sviridova, S.N.Yudin\thanks{e-mail:yudin@helene.sinp.msu.ru}}
\date{}
\maketitle

\centerline{\small{Institute of Nuclear Physics, Moscow State University
119899 Moscow, Russia}}

\begin{abstract}
The nonlinear model of pionic condensate in nuclei by G. Preparata can 
be efficiently verified by investigation of the quasielastic knockout 
process $A(e,e' \pi^\pm )A^*$. First, a momentum distribution (MD) of the 
collective pions has a bright maximum at $q_0 \simeq 0.3$ GeV. Second,
the excitation spectrum of a recoil nucleus is concentrated at low 
energies $E_{rec,A} \simeq q_0^2/2Am_N \le 1$ MeV. The results for the 
pion knockout from mesonic clouds of individual nucleons are absolutely 
different. The latter results are presented both for pion and 
$\rho$-meson clouds localized on nucleons.
\end{abstract}

\section{Introduction}

For the last decade the problem of mesons (in particular, pions) in cold and hot nuclei
has been one of the most urgent problems of nuclear and hadron physics ~\cite{1a,1b,1c}.
Is the number of pions per nucleon in nuclei bigger than in a free nucleon?
Do the properties of mesons in nuclei change as compared with ones in vacuum?
Do pions give essential additional contribution to the nuclear quark condensate?
This is not a full list of urgent questions.

One of the most interesting questions in this field is the question about 
a possibility of separating pions from nucleons in nuclei and pions 
forming a collective degree of freedom. In the 70's, A.B. Migdal offered 
a concept of a pion condensated state in nuclei ~\cite{1} (for details, see 
the monograph ~\cite{2}). Here, three coupled channels in nucleus are 
considered: a nucleon on an excited shell-model orbital and a nucleon hole, 
$\Delta$-isobare and a nucleon hole and pions.
According to later estimates ~\cite{3}, the real nucleon density in nuclei
is less than critical one, which is necessary to form a condensate.
In the 90's, G. Preparata with his colleagues ~\cite{4} proposed a new 
mechanism of pionic injecting into nuclei. He followed ideas by R. Dicke about the 
supperradiation in a system of many identical atoms ~\cite{5}, i.e. about the
coherent photon emission. The power of the emission in this system can be
$N^2/4$ times stronger than that in an isolated atom ($N$ is the number
of atoms). It means that a "superstrong" interaction of a photon field with the
system of atoms appears.

G.Preparata extended this concept to a system of nucleons, where $N$ and 
$\Delta$-isobare play the role of the ground state and an excited state of 
an atom, respectively, while the pion field is analogous to the photon one.
According to the estimates by G. Preparata, in spite of the large energy losses
($\sim$ 300 MeV), which are necessary to excite the $\Delta$-isobare, a
powerful coherent interaction between $N$-, $\pi$- and $\Delta$ subsystems 
results in pion condensation in nuclei even at the real density $\rho_{nucl.}$.
This is accompanied by the remarkable increase of the binding energy of the 
nucleus (which is equal to $\sim$ 60 MeV per nucleon). The coherence domain
includes approximately 70 nucleons. This "superradiative" pionic mode is
characterized by momenta of pions $\vec{q}_0$,

\begin{equation}
\sqrt{\vec{q}_0^2+m_\pi^2}=m_\Delta-m_N, \qquad q_0 \simeq 0.3 \quad {\rm GeV},
\end{equation}
$m_\Delta$ and $m_N$ are masses of $\Delta$ and $N$.

The number of the collective pions $n_{i,coll}$ of each kind $i$ ($\pi^+$, $\pi^-$
or $\pi^0$) per nucleon in nuclei with $N=Z$ is approximately  0.1
~\cite{4}. It is larger than the corresponding amount of pions in the cloud
of an individual nucleon having momenta within $k=(1 \pm 0.1)q_0$ range
(see below, we mean the $P$-state of the pion in the channel of virtual decay
$p \to n+\pi^+$ ~\cite{6,7}).

The experimental situation with pions in nuclei is contradictory. 
For a long time it
 was thought that in nuclei there is a definite deficit 
of pions. In any case, the number of pions in nuclei is smaller that the 
RPA approach predicts. This was reflected in the title of the wellknown paper 
"Where are the nuclear pions ?" \cite{7a}.
Recently, however an enhancement of pionic degrees of freedom predicted by 
RPA was observed experimentally. Namely, the longitudinal nuclear response 
$R_L$ in the ($\vec{p} \vec{n}$)-reaction on $^{12}C$, $^{40}Ca$, $^{28}Pb$ 
nuclei was measured \cite{8}. This reaction is characterized by transfer
of pion quantum numbers from proton to the nucleus. Values of $R_L$ at the momentum transfer 
$q=1.7$ fm$^{-1}$ and energy transfer $\omega$ around 60 MeV indicate derectly
the pionic degrees of freedom in nuclei (for the $^{12}C$ 
target the experiment showed an interesting collective enhancement, indicating 
excitation of the corresponding charged giant resonance ~\cite{8}).  
Now, the interest in the  problem of pion content of nuclei revives. 
It is necessary to point out that "Preparata pions" could not probably be observed as an 
additional longitudinal responce in $(p,n)$-reactions. It is related to an 
important role of the $\Delta$-isobares in the hypothetical Preparata phenomenon. 
In the usual consideration of the relation between nucleon particle-hole 
responce and contents of pions the role of $\Delta$-isobares is not taken 
into account.

The main aim of the present work is to propose a very independent kind of 
experiment, which gives us a possibility to see "Preparata pions", i.e.to see
directly a momentum distribution (MD, the square of a wave function in the momentum 
representation) of collective pions, which are not localized on the nucleon. We 
mean the reaction of quasielastic knockout (QEK) of pions from the nucleus by 
electrons with energy of a few GeV.

The background of QEK in microphysics is very rich.
The exclusive processes of the quasielastic knockout of protons
from the atomic
nucleus by protons ($p,2p$) or by electrons ($e,ep$) at bombarding energies
of a few hundred MeV  are well known ~\cite{9}. they were 
used for  investigations of  the MDs of nucleons on separated shell-model 
orbitals. The shape of the MD for light nuclei was proved out to be  very 
sensitive to values
of the nucleon shell-model quantum numbers $nl$.
  
Experimentally, the coincidence technique is used with the energy resolution 
$\Delta E \sim 1$ MeV, and the interpretation of results  is based on a very 
simple binary conservation laws

\begin{equation}
E_0=E_1+E_2+E_{bind},
\qquad
\vec{p}_0+ \vec{q}= \vec{p}_1+ \vec{p}_2,
\end{equation}
which are valid here, because energies of both initial  bombarding
particles and two final particles are high (the impulse approximation ~\cite{10}).
In Eq.2, $\vec{q}$ is the initial momentum of a virtual particle to
be knocked out, and the rest of the notations is evident. The kinematics of the
above process corresponds to the inequalities $|\vec{q}| \ll |\vec{p}_0|, |\vec{p}_1|,
|\vec{p}_2|$ and $E_{bind} \ll E_0, E_1, E_2$. In this way, the values of
$\vec{q}$ and $E_{bind}$ are obtained from the experiment.

One more nuclear example is exclusive reactions ($p,p \alpha)$, etc. of 
the quasielastic knockout of 
nucleon clusters by protons with the energy 0.5-1 GeV. This reaction can be
important for identification at high energies ~\cite{11} the  mechanisms of 
deexcitation of virtually excited clusters in nuclei.

The QEK process $(e,2e)$ at the beam energies of around 10 KeV with the 
analogous extraction of the electronic MDs is widely used in investigations of 
the electronic structure of atoms, molecules and solids ~\cite{12,13}. So, there 
is a great experience in the investigation of the exclusive QEK reactions.
In our previous papers ~\cite{6,7,14} we have extended  the concept of the 
QEK to the knockout of  mesons 
from nucleons by high energy electrons. It demanded a relativistic
generalization of the theory. Namely, the pole $z$-diagram reflecting a virtual 
creation of, say, $\pi^+ \pi^-$ pair was taken into account  in addition to the 
usual diagram of pion knockout (the instantaneous form of dynamics). The second 
important point was that it is possible to separate experimentally the 
reactions induced by longitudinal virtual photons $\gamma_L^*$ and ones induced by
transverse virtual photons $\gamma_T^*$ ~\cite{15}. This offers a unique way
~\cite{16} to investigate  the MDs of pions ($\pi^{+*}+\gamma^*_L \to
\pi^+$ subprocess) and $\rho$-mesons ($\rho^{+*}+\gamma^*_T \to \pi^+$ 
subprocess) separately, by means of the $p(e,e' \pi^+)n$ reaction with 
the squared mass of virtual photon $Q^2$ of about 2-4 (GeV/$c$)$^2$.

In the present paper we extend this approach to the investigation of the pionic 
degrees of freedom in nuclei. The paper is organized as follows. In the second 
section we represent shortly a relativistic formalism for the QEK reactions.
In the third section, a simple expression for the MDs of the collective pion 
in the 
Preparata model is derived and compared with that of pions localized on
nucleons in a nucleus. In the fourth section, another variable, which is 
observable in 
the $(e,e' \pi)$ experiment, the MD of $\rho$-mesons in the nucleus is 
calculated within the model of $\rho$-meson localization on the nucleons in a
nucleus. The fifth section outlines the advantage of the $(e,e' \pi)$
process in comparison with a $(\gamma, \pi)$ reaction and $(\pi, 2\pi)$ QEK
process.

\section{Formalism}

According to the general theory of meson electroproduction ~\cite{17}, the 
differential cross section of the reaction $T+e \to R+\pi^++e'$ may be
separated into longitudinal (L) and transverse (T) components, along with
interference terms by varying kinematical variables of the
final detected particles, $\pi^+$ and $e'$ (Rosenbluth separation). This 
important result is usually discussed in terms of the independent variables
$Q^2=-q^2$ ($q$ is a 4-momentum of the virtual photon), $W^2=(p_\pi+p_B)^2$
($W$ is the invariant mass of the final hadrons), $t=(p_B-p_p)^2$ ~\cite{15,17}.
Such a parametrisation is convenient, if we have in mind a broad 
kinematical region, including, say, resonances. But in a narrow region 
corresponding to the quasielastik knockout, the traditional five variables 
$E_e'$, $\Omega_{e'}$, $\Omega_\pi$ of the nonrelativistic QEK theory are
more efficient~\cite{6,7}.

Having in mind this remark, we can write  for the longitudinal 
cross section of the QEK process on a nucleus an expression, that is very
close to formulae in ~\cite{6,7}

\begin{equation}
\frac{d^5 \sigma_L(eT \to e' \pi R)}{dE_{e'}d \Omega_{e'}d \Omega_\pi}=
\frac{E_{e'}^2 }{E_e} \frac{I(e \pi)}{E_\pi(\vec{k}')}
4 \overline{|\Psi^{R \pi}_T(\vec{k})|^2}
\frac{d \sigma(e \pi \to e' \pi)}{d \Omega_\pi},
\end{equation}
where the right part contains only one, predominant pole component 
~\cite{7,8,14} at $Q^2 \approx 2-3$ (GeV/$c$)$^2$. Here, $E_{e'}$ and $\Omega_{e'}$ 
are  energy
and a solid angle of the final electron; d$\Omega_\pi$ is an element of a solid
angle of the final pion in the lab system; $E_{e}$ is  
energy of the initial electron, $\vec{k}$ is momentum of the virtual pion;
$E_\pi(\vec{k}')=\sqrt{\vec{k}'^2+m_\pi^2}$; $\vec{k'}$ and $E_\pi(\vec{k}')$ are
momentum and energy of the knocked-out real pion; 
$I(e \pi)$ is the invariant flux of electrons and pions;
$\overline{|\Psi^{R \pi}_T(\vec{k})|^2}$ is the MD of pions in the channel
$T \to R\pi$; and, finally, $d \sigma (e \pi \to e' \pi)/d \Omega_\pi$ is a cross 
section of a free scattering process $e+\pi^+ \to e'+\pi'^+$. The bar means spin 
projection average quantity.

The transverse cross section is given by a formula very close to the
formulae in~\cite{7,14}

\begin{equation}
\frac{d^5 \sigma_T}{dE_{e'}d \Omega_{e'}d \Omega_\pi}=
\frac{E_{e'}^2 }{E_e} \frac{I(e \rho)}{E_\rho(\vec{k})}
\overline{|\Psi^{B \rho}_p(\vec{k})|^2}
\frac{d \sigma(e \rho \to e' \pi)}{d \Omega_\pi}.
\end{equation}
Here $E_\rho(\vec{k})=\sqrt{\vec{k}^2+m_\rho^2}$.
Eq.(4) is valid at
$Q^2 > 2$ (GeV/$c$)$^2$ ~\cite{14}.

Both free cross sections include   relevant electromagnetic form factors 
~\cite{7,14}.

The differential cross sections $d \sigma /d \Omega_\pi$ for the both mechanisms 
are given by

\begin{equation}
\frac{d \sigma}{d \Omega_\pi}=
\frac{\overline{|M_{fi}|^2}}{64 \pi^2}
\frac{|\vec{k}'|}{m_{\pi,\rho}E_e^2}
\frac{1}{1-(E_\pi(\vec{k}')/|\vec{k}'|)cos \theta_{\pi' e}},
\end{equation}
where $\theta_{\pi' e}$ is an angle between the incident electron and the final pion.

For the elastic scattering off point pions
the square amplitude is equal to

\begin{equation}
\overline{|M_{fi}|^2}=
64 \pi^2 \frac{m_\pi E_e}{E_{e'}} \sigma_M;
\end{equation}
for the scattering off point $\rho$-mesons (with de-excitation) (see Appendix)

\begin{equation}
\overline{|M_{fi}|^2}=
64 \pi^2 \frac{g_{\rho \pi \gamma}^2}{m_\pi^2}
\frac{E_e}{E_{e'}} \sigma_M
\left[tg^2(\theta /2) \left( (lk)(kq)-3/4m_\rho^2Q^2 \right)-1/2(kq)^2
\right],
\end{equation}
where $g_{\rho \pi \gamma}$ is the constant of conversion of the $\rho$-meson
into pion, $\sigma_M$ is the Mott cross section 
$\sigma_M=(4 \alpha^2cos^2(\theta /2)E_{e'}^2)/Q^4$, $\theta$ is an angle 
between the incident and scattered electrons, $l$, $k$, $q$ are 4-momenta of the
incident electron, $\rho$-meson and photon.

\section{Momentum Distributions of delocalized pions in nuclei}

Having in mind the Preparata model ~\cite{4}, we suppose for simplicity, that 
the coherence domain coresponds to the whole nucleus. Hence, a radial wave 
function of the collective pion has a form of the standing $P$-wave ~\cite{4}

\begin{equation}
\Phi (r)=cj_1(q_0r), \qquad r \le R
,
\end{equation}
$$
\Phi (r)=0, \qquad r > R,
$$
and nucleons in the initial and final nuclear states occupy the same 
shell in the mshell-model, transitions like $0^+ \to 1^+$, $1^+ \to 1^+$, 
etc. take place. The situation is close to weak coupling of the pion to 
the nucleus. In Eq. (8), $R$ means the radius of the domain. Its value 
should correspond to $A \simeq 70$ at $N=Z$. Next, the constant $c$ is 
defined by the normalization of the wave function (8) to the 
abovementioned value $n_{i,coll}=0.1$, so $c=0.027$. By the Fourier
transformation of the single-particle wave function (8), we obtain the MD
of the collective pions $\overline{|\Psi_A^{A \pi}(\vec{k})|^2}$. 
This MD corresponds to the longitudinal cross
section (3), (5) and is presented in Fig.1 (solid curve) with normalization 
to one nucleon to facilitate its comparison with the dashed curve 
(the MD of pions in a free nucleon, see below). The solid curve should be 
multiplied by $A$ to compare it with experimental data. It has a bright 
maximum at $k=q_0$, i.e. it is close to the plane wave within the 
limitation imposed by the finite volume of the nucleus. The dashed 
curve corresponds to the MD of the pion in a free nucleon 
$\overline{|\Psi_N^{N \pi}(\vec{k})|^2}$ ($N \to N+\pi$ channel of the
virtual decay), which we have reconstructed ~\cite{6,7} from the $p(e,e' \pi)n$
experiment ~\cite{15}. This experiment, in fact, was carried out at the
quasielastic kinematics ~\cite{6,7,14} ($Q^2$ was large enough: 
$Q^2=1-3$ (GeV/$c$)$^2$). Our analysis ~\cite{6,7,14} was based on the 
relativistic pole approximation in the laboratory system, which 
included a pole $z$-diagram.

Fig.2 represents a washed-out MD of the localized pions
$\overline{\overline{|\Psi_N^{N \pi}(\vec{k})|^2}}$ (thick solid line) 
obtained by the convolution of
the MD of pions in a free nucleon (dashed line) with the averaged MD of 
the shell-model nucleons in nucleus 
$\overline{|\Phi_A^{A-1,N}(\vec{p})|^2}$ (solid line):

\begin{equation}
\overline{\overline{|\Psi_N^{N \pi}(\vec{k})|^2}}=
\int \overline{|\Psi_N^{N \pi}(\vec{k}+(m_\pi/m_N) \vec{p})|^2}
\cdot \overline{|\Phi_A^{A-1,N}(\vec{p})|^2}d \vec{p}.
\end{equation}

Both the MD of pions in a free nucleons and the washed-out MD of the localized pions, 
in contrast to the MD of pions in the Preparata model, are
very smooth at $k$ values, which are close to $q_0$. The momentum 
distributions corresponding to Fig.1,2 are isotropic with respect to $\vec{q}$
direction, because they are averaged over the magnetic quantum numbers of the
pionic $P$-orbital.

The main point here is that the knockout of the 
delocalized collective pions is accompanied by recoil to the final
nucleus as a whole, with the corresponding very small recoil energy
$E_{rec,A} \simeq q_0^2/2Am_N <1$ MeV ($A \simeq 70-80$), although the momentum
$q_0$ itself is not small (see Eq. 1). This is why such prediction is not
trivial. The MD of such pions has a sharp maximum at $k=q_0$.

The wave function of pions (8) does not contain pion-nucleon spacious
correlations and, as a result, the final recoil nucleus will not be internally
excited. In fact, the best experimental energy resolution $\Delta E$ may be
around 10 MeV here, and this severe simplification partly softens. The real
situation will correspond to the summation over many exited states of the external
shell of the final nucleus, i.e. to a sum rule.

At the same time, the opposite extreme case of the knockout of pions with the
same virtual momentum $\vec{k}$, $k \simeq q_0$ from the pion cloud
of an individual nucleon will be characterized by a large value of the recoil 
energy transfered to one nucleon, $E_{rec,N} \simeq q_0^2/2m_N \simeq 50$ MeV. 
This nucleon, with a large probability, will be directly emitted from the
nucleus (we mean here the numerous weakly bound nucleons of the external shell).
A reliable identification of this event requires triple coincidences
$e'+\pi^-+p$, which is an urgent experimental problem. In a noncomplete
experiment with only double coincidences $e'+\pi^-$, the abovementioned event 
will be perceived as
a corresponding high excitation ($\omega \simeq 50$ MeV) of the final recoil 
nucleus accompanied by the transfer of the momentum $-\vec{k}$, $k \simeq q_0$ to this
nucleus. This group of events will show a very smooth MD of the virtual pions
like the corresponding part of the thick solid line in Fig.2 at $k$ around $q_0$.
The discussed recoil of the proton in the process of the $\pi^-$ knockout 
from the nucleus can also create one of the charged nuclear giant resonances
with the excitation energy $\omega$ of a few tens MeV (see above, the discussion
of the $(\vec{p}, \vec{n})$ meson transfer reaction). An intriguing new 
opportunity here is to investigate a $k$-dependence of such cross sections.

So, the principal result of this section is that the high-energy pion 
electroproduction on nuclei by means of the virtual longitudinal photons 
$\gamma_L^*$ at the kinematics of the QEK process at small $\omega$ values
of a few MeV offers an opportunity to see the cooperative, maximally
delocalized pions in nuclei by the most direct way. The bright 
MD maximum at $k \simeq q_0 \simeq 0.3$ GeV/$c$ (the solid line in Fig.1) will
be the principal sign of the presence of such pions in the nucleus. The 
increase of $\omega$ corresponds, qualitatively, to increasing localization
of the discussed virtual pions in the nucleus, and the shape of the MD measured
at different $\omega$ may be helpful for the clarification of evolution of 
the reaction mechanism with the increase of the excitation energy $\omega$ (the 
usage of the triple coincidence would be very urgent here, too).

It must be noted, that final energy of the knocked-out pion should be not lower 
than 1 GeV to avoid disturbing influence of the intermediate resonance 
($\Delta$-isobare) in the $\pi-A$ final state interaction. Energies much higher
than 1 GeV are also not suitable because they correspond to the different 
physics of asymptotical quark counting rules ~\cite{20}.
 This physics, which corresponds 
to the $Q^2$ values of 10-20 (GeV/$c$)$^2$ and rather small cross sections, is very popular now
~\cite{20a}. But our physics of the soft hadronic degrees of freedom in the nucleons and 
nuclei, which corresponds to rather moderate $Q^2$ values of 2-4 (GeV/$c$)$^2$ and which is 
unfortunately still in the shadow, is not less interesting. By the way, it corresponds to the 
quite measurable cross sections.

\section{Momentum distributions of $\rho$-mesons localized on nucleons in 
nuclei}

As it was noted in the section 1, here we will discuss the model of $\rho$-mesons 
localized on nucleons in the nucleus. The MD of the $\rho$-mesons may
be obtained by means of the QEK reaction such as 
$A(N,Z)(e,e' \pi)A(N-1,Z+1; \omega)$ initiated by the transverse virtual photons,
$\rho^-+\gamma_T^* \to \pi^-$. The principal experimentally observed difference
of the model under consideration from the discussed above model of the pionic 
degrees of freedom is that the MD should be the same for any $\omega$ value,
i.e. that the MD does not depend on a microscopic process, which follows 
the recoil to the nucleon (at the small $k$, $k \le 0.1$ GeV/$c$, either a soft
nuclear excitation within the lowest shell-model configuration for small
$\omega$ of a few MeV, or an excitation of the charged giant resonance for
$\omega \simeq 20-40$ MeV take place; at the relatively large $k$, $k \simeq q_0$,
either an excitation of the charged giant resonance at $\omega \simeq 30$ MeV 
or a direct emission of the recoil nucleon at $\omega \simeq 50-60$ MeV take 
place, etc.).

Starting with the $\rho$-meson MD in the nucleon 
$\overline{|\Psi_N^{N \rho}(\vec{k})|^2}$ (dashed line in Fig.3) ~\cite{7,14},
we take into account Fermi motion of nucleons in the nucleus by means of Eq.(9)
and obtain, finally, the solid curve in Fig.3 analogous to the
thick solid curve in Fig.2 but extended to much larger values of $k$ due to
the large value of the mass $m_\rho \simeq 800$ MeV. This curve represents the
washed-out MD of $\rho$-mesons in the nucleus 
$\overline{\overline{|\Psi_N^{N \rho}(\vec{k})|^2}}$
if they are localized on nucleons. 
A deviation of experimental results from this shape, similar to the deviation 
of the solid curve in Fig.1 from the thick solid curve in Fig.2 will mean the 
delocalization of $\rho$-mesons in the nucleus (see above, section 2).

\section{Conclusion}
In this paper, we have proposed a program of the direct experimental
investigation of the pionic wave functions in nuclei by means of the quasielastic
knockout reactions $A(N,Z)(e,e' \pi^-)A(N-1,Z+1; \omega)$ or
$A(N,Z)(e,e' \pi^+)A(N+1,Z-1; \omega)$ initiated by electrons with energy of 
a few GeV and mediated by the longitudinal virtual photons. It has been 
demonstrated, that the momentum distribution of the delocalized pions in 
the cooperative model by Preparata is qualitatively different from the MD of 
the pions localized on nucleons in the nucleus. It is expected that the 
corresponding spectra of excitation energies of the final nucleus-spectator, 
$\omega$, will be rather different in these two cases.

Futher, having in mind the $(e,e' \pi^-)$ or $(e,e' \pi^+)$ reaction mediated by
the transverse photons, we have demonstrated, as a basic example, the MD of 
$\rho$-mesons in nucleus within the simplest model of $\rho$-mesons localized 
on nucleons in the nucleus.

Recently, the investigation of the $(\pi, 2 \pi)$ quasielastic knockout process
has been initiated ~\cite{21}, although the energies of the knocked-out pions 
are still not high enough. The cross sections here are larger (strong 
interaction) than those for the $(e,e' \pi)$ reaction, and the $(\pi, 2 \pi)$
reaction gives an opportunity to investigate the $\pi^0$-component of the
collective field,  but distortion and absorption effects ~\cite{22} for
three pionic waves in the $(\pi, 2 \pi)$ reaction are much more pronounced 
than those for one pionic wave in the $(e,e' \pi)$ process. So, these two
reactions, the $(e,e' \pi)$ reaction of
a volume character with smaller cross sections and the $(\pi, 2 \pi)$ reaction
of a surface character with larger cross sections, can complement each other 
rather efficiently.

Finally, it should be noted, that the analogous process $( \gamma, \pi)$ on 
nuclei ($Q^2=0$) corresponds to the interference of amplitudes for a few 
different diagrams ~\cite{23} and does not offer a direct way for extracting 
the MDs of pions in nuclei.

\bigskip

The authors are grateful to Profs. A.A. Ogloblin and B.S. Slowinsky for their
lively interest in the discussed problems.

\bigskip

This work is supported by the Russian Foundation of Fundamental Research
(grant N 00-02-16117).

\newpage
\section{Appendix}

We will present here the main steps of calculating the amplitude
of the process
 $e+ \rho \to e' + \pi$. 

The invariant amplitude $M_{fi}$ for this process is equal to

\begin{equation}
M_{fi}= \frac{e^2g_{\rho \pi \gamma}}{Q^2m_\pi}(\bar{u'} \gamma_\beta u)
\varepsilon^{\beta \mu \alpha \nu}e_{m \mu}q_\alpha k_\nu .
\end{equation}

Here $g_{\rho \pi \gamma}$ is the constant of $\rho \pi \gamma$-interaction,
$\bar{u'}$, $u$ are the Dirac's spinors of the electron (with the normalization 
$\bar{u}u=2m_e$), $\gamma_\beta$ are the Dirac's matrices,
$\varepsilon^{\beta \mu \alpha \nu}$ is a an antisimmetrical unity tensor,
$e_{m \mu}$ is a vector of polarisation of the $\rho$-meson.

The summarizing over polarizations should be clarified.

\begin{equation}
\overline{|\bar{u'} \gamma_\beta u \cdot
\varepsilon^{\beta \mu \alpha \nu}e_{m \mu}q_\alpha k_\nu|^2}= \frac{1}{2}
 \frac{1}{3}
\sum_{m \mu \mu '}(\bar{u'} \gamma_\beta u)(\bar{u'} \gamma_{\beta '} u)^*
\varepsilon^{\beta \mu \alpha \nu}e_{m \mu}q_\alpha k_\nu
\varepsilon^{\beta ' \mu ' \alpha ' \nu '}e_{m \mu '}^*q_{\alpha '} k_{\nu '}.
\end{equation}

Having calculated electron traces, we obtain

\begin{equation}
\overline{|\bar{u'} \gamma_\beta u \cdot
\varepsilon^{\beta \mu \alpha \nu}e_{m \mu}q_\alpha k_\nu|^2}= 
\frac{1}{3}
2 (2l_\beta l_{\beta '}-g_{\beta \beta'}(ll'))
\varepsilon^{\beta \mu \alpha \nu}
\varepsilon^{\beta ' \mu ' \alpha ' \nu '} 
\left( -g_{\mu mu'} +\frac{k_\mu k_{\mu'}}{m_\rho ^2} \right)
q_{\alpha} k_{\nu}q_{\alpha'} k_{\nu'}
\end{equation}

$$
=\frac{2}{3}(2l_\beta l_{\beta'} \varepsilon^{\beta \mu \alpha \nu}
\varepsilon^{\beta ' \alpha ' \nu '}_\mu 
q_{\alpha} k_{\nu}q_{\alpha'} k_{\nu'}
-(ll') \varepsilon^{\beta \mu \alpha \nu}
\varepsilon^{\alpha' \nu'}_{\beta \mu} 
q_{\alpha} k_{\nu}q_{\alpha'} k_{\nu'}).
$$

The unity tensors are convoluted according to the following rules:

\begin{equation}
\varepsilon^{\beta \mu \alpha \nu} \varepsilon^{\alpha' \nu'}_{\beta \mu} =
=2 \varepsilon^{\alpha \nu}_{\alpha' \nu'}= 
2[\delta_{\alpha \alpha'} \delta_{\nu \nu'} -
\delta_{\alpha \nu'} \delta_{\alpha' \nu}]
\end{equation}

\begin{equation}
\varepsilon^{\beta \mu \alpha \nu}
\varepsilon^{\beta ' \alpha ' \nu '}_\mu =
\varepsilon^{\beta \alpha \nu}_{\beta ' \alpha ' \nu '}.
\end{equation}

Thus, we obtain

\begin{equation}
\overline{|\bar{u'} \gamma_\beta u \cdot
\varepsilon^{\beta \mu \alpha \nu}e_{m \mu}q_\alpha k_\nu|^2}= 
4 [(ll')(q^2k^2-(kq)^2)-2(lk)(lq)(kq)+(lq)^2k^2+(kl)^2q^2].
\end{equation}

This equation permits us to find easily the cross section (7).

\newpage
\begin{figure}[h]
\epsfverbosetrue
\epsfysize=15cm
\epsfxsize=15cm
\epsfbox{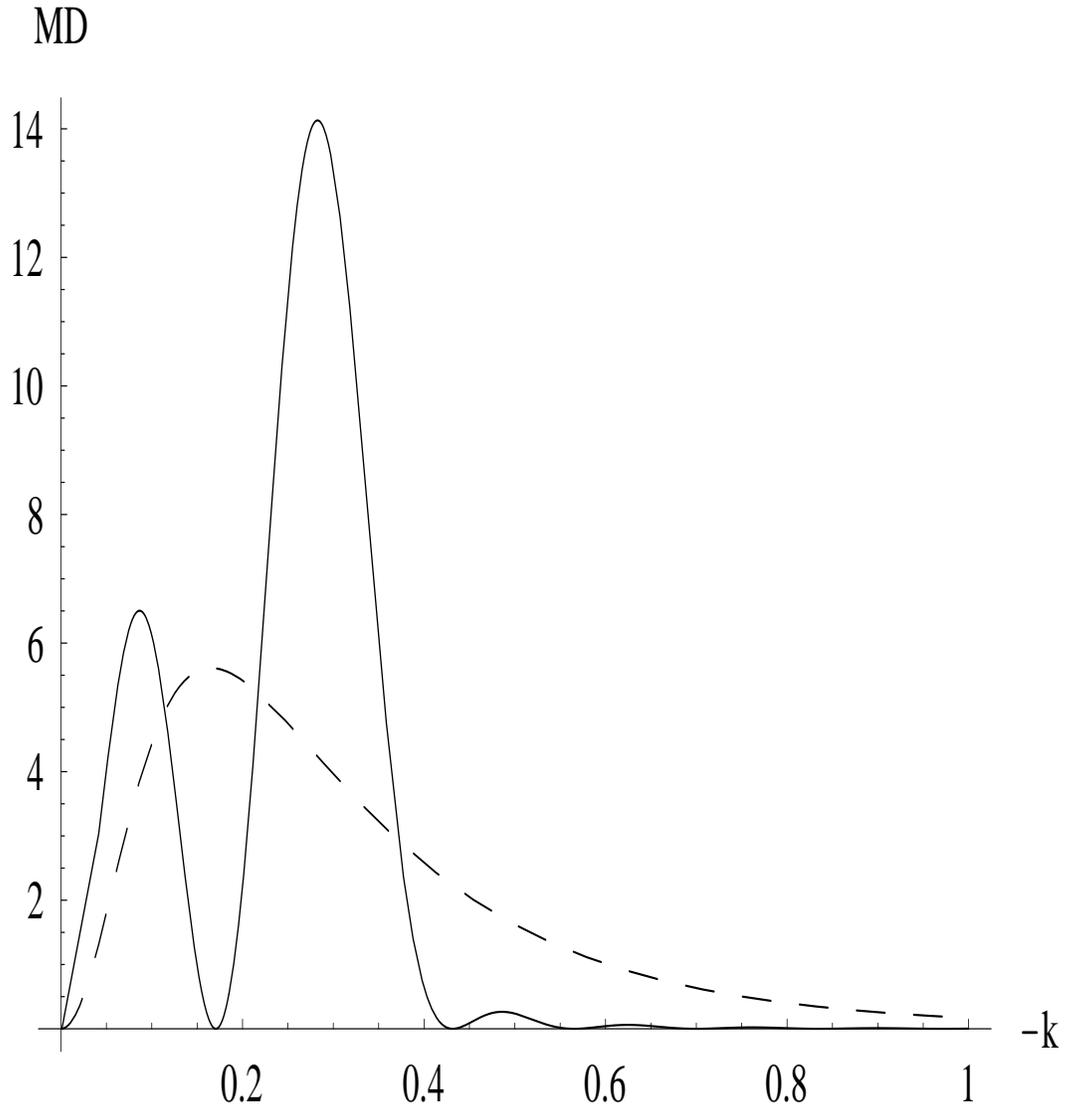}
\caption{Momentum distribution of pions, versus $|k|$, GeV/c:
solid curve - the MD of pions in nuclei, Preparata model 
$\overline{|\Psi_A^{A \pi}(\vec{k})|^2}$, (GeV/c)$^{-3}$;
dashed curve - the MD of pions in a free nucleon 
$\overline{|\Psi_N^{N \pi}(\vec{k})|^2}$, (GeV/c)$^{-3}$.}
\end{figure}

\newpage
\begin{figure}[h]
\epsfverbosetrue
\epsfysize=15cm
\epsfxsize=15cm
\epsfbox{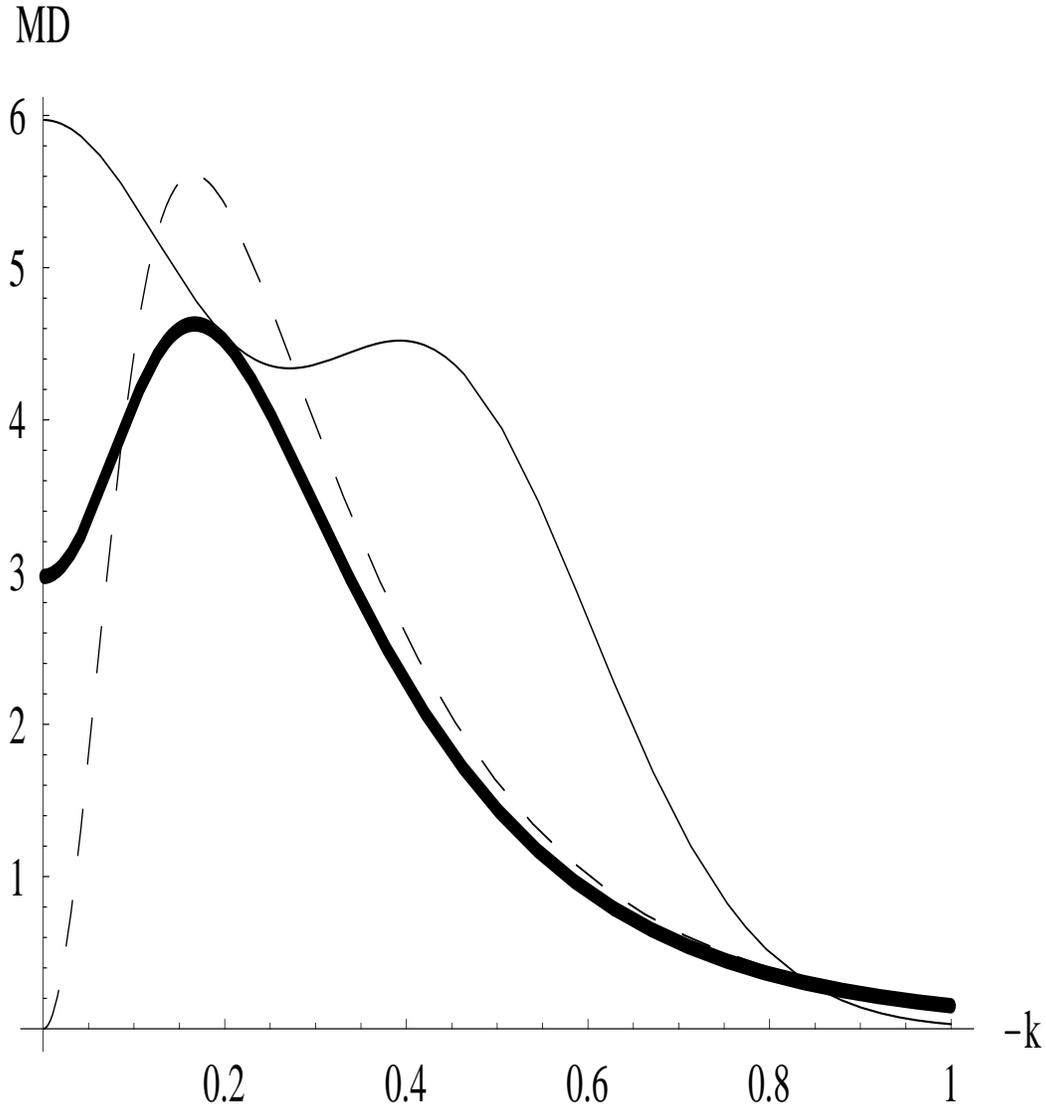}
\caption{MD of pions in a nucleus versus $|k|$, GeV/c. 
Dashed line is the MD of pions in a free nucleon
$\overline{|\Psi_N^{N \pi}(\vec{k})|^2}$, (GeV/$c$)$^{-3}$.
Thin solid line is the average MD of a nucleon in the nucleus
$\overline{|\Phi_A^{A-1,N}(\vec{k})|^2}$, (GeV/$c$)$^{-3}$.
Thick solid line is the washed-out MD of localized pions in the nucleus
$\overline{\overline{|\Psi_N^{N \pi}(\vec{k})|^2}}$, (GeV/$c$)$^{-3}$.}
\end{figure}

\newpage
\begin{figure}[h]
\epsfverbosetrue
\epsfysize=15cm
\epsfxsize=15cm
\epsfbox{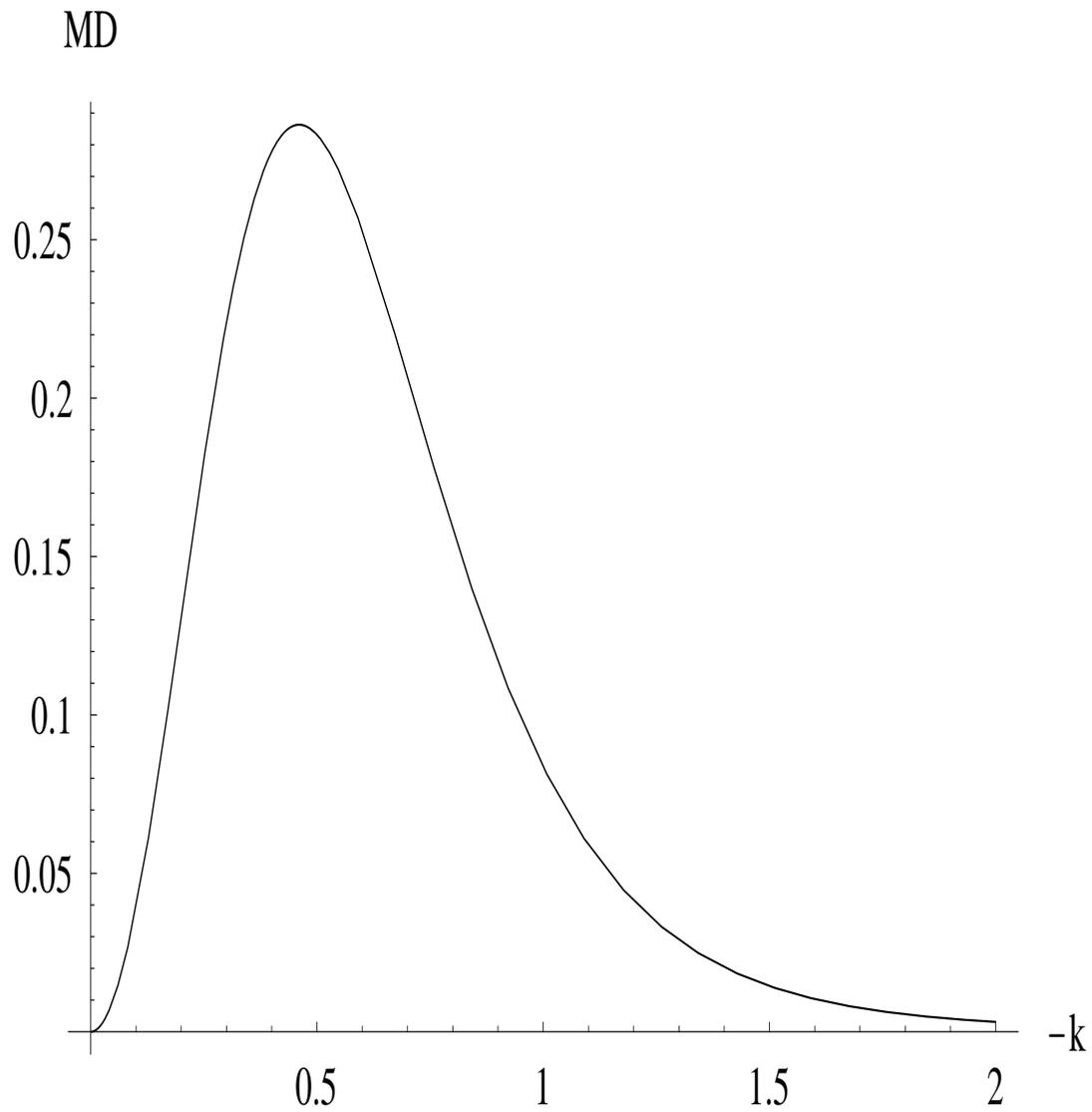}
\caption{MD of the $\rho$-mesons in a nucleus versus $|k|$, GeV/c.
Dashed line is the MD of the $\rho$-mesons in a free nucleon
$\overline{|\Psi_N^{N \rho}(\vec{k})|^2}$, (GeV/$c$)$^{-3}$.
Solid line is the MD of the $\rho$-mesons in a nucleus
$\overline{\overline{|\Psi_N^{N \rho}(\vec{k})|^2}}$, (GeV/$c$)$^{-3}$.}
\end{figure}

\end{document}